\documentclass[note]{aa}
\pdfoutput=1
\usepackage[utf8x]{inputenc}
\usepackage{natbib}
\usepackage{amsmath,txfonts,graphicx,varioref}
\makeatletter
\def\includeplaatje#1{\includegraphics[width=.95\columnwidth]{#1.eps}}%
\ifx\pdfoutput\@undefined\else
  \ifx\pdfoutput\relax\else
    \ifcase\pdfoutput\else
      \global\def\includeplaatje#1{\includegraphics[width=.95\columnwidth]{#1.pdf}}%
\fi\fi\fi
\makeatother
\def\lofar{Lofar}

\def\aph{Astropart.\ Phys.}

\def\jetp{Soviet Phys. \textsc{jetp}}
\def\dans{Dokl. Akad. Nauk. \textsc{sssr}}
\def\nimpa{Nucl. Instrum. Meth.~A}
\def\GZ{Ge\-ra\-si\-mova\discretionary{–}{}{–}Za\-tse\-pin}
\def\e#1{\cdot10^{#1}}
\begin{document}
% *****************************************************************************************************
\title{Prospects for direct cosmic ray mass measurements through the \GZ\ effect}
\date{April 3, 2008}

\author{S.~Lafèbre\inst{1}
		\and
	H.~Falcke\inst{1,2}
		\and
	J.~Hörandel\inst{1}
		\and
	J.~Kuijpers\inst{1}}

\offprints{S.~Laf\`ebre (s.lafebre@astro.ru.nl)}

\institute{Department of Astrophysics, IMAPP, Radboud University, P.O.~Box 9010, 6500GL Nijmegen, The Netherlands\\
             \and
           Radio Observatory, Astron, Dwingeloo, P.O.~Box 2, 7990AA Dwingeloo, The Netherlands}

\date{Received $\langle$date$\rangle$ / Accepted $\langle$date$\rangle$}

\abstract{%
The Solar radiation field may break apart ultra high energy cosmic nuclei, after which both remnants will be deflected in the interplanetary magnetic field in different ways. This process is known as the \GZ\ effect after its discoverers.
}{
We investigate the possibility of using the detection of the separated air showers produced by a pair of remnant particles as a way to identify the species of the original cosmic ray primary directly. Event rates for current and proposed detectors are estimated, and requirements are defined for ideal detectors of this phenomenon.
}{
Detailed computational models of the disintegration and deflection processes for a wide range of cosmic ray primaries in the energy range of $10^{16}$ to~$10^{20}$~eV are combined with sophisticated detector models to calculate realistic detection rates.
}{
The fraction of \GZ\ events is found to be of the order of $10^{-5}$ of the cosmic ray flux, implying an intrinsic event rate of around $0.07$~km$^{-2}$\,sr$^{-1}$\,yr$^{-1}$ in the energy range defined. Event rates in any real experiment, however, existing or under construction, will probably not exceed $10^{-2}$~yr$^{-1}$.
}{}


\maketitle
% *****************************************************************************************************
\section{Introduction}
The mass composition of very high energy cosmic rays provides important information on their acceleration mechanisms and the compositions of their sources. Usually, however, it is only possible to make statistical, model-dependent estimates of the primary particle types from an ensemble of showers. Primary compositions are then derived from the abundances of species components in the air showers considered~\citep{2002:Kascade}.

An alternative mass determination makes use of the \emph{\GZ\ effect}~\citep{1951:Zatsepin,1960:Gerasimova}. In this scenario, one relies on the fact that atomic nuclei have a chance of undergoing photodisintegration in the Lorentz boosted Solar radiation field before arriving at Earth, splitting the nucleus into two parts. Due to the different rigidities of these two fragments, the deflection in the interplanetary magnetic field will be different, resulting in two separate air showers at some spatial separation, but arriving essentially at the same time and from the same direction \citep{1999:Medina-Tanco,1999:Epele}. To our knowledge, no experimental detection of a \GZ\ event has ever been reported.

% *****************************************************************************************************
\section{The \GZ\ process}\label{sec:proces}

The photodisintegration probability~$\eta_Z$ for a nucleus approaching the Solar system to undergo photodisintegration has been investigated thoroughly by \cite{1951:Zatsepin}, \citet{1960:Gerasimova}, \citet{1999:Medina-Tanco} and \citet{1999:Epele}. It can be calculated by integrating its path length against photodisintegration over its trajectory. The photon energy as seen from the cosmic ray's comoving frame is Lorentz boosted by a factor $2\gamma\cos^2(\alpha/2)$, where $\gamma$~is the cosmic ray's Lorentz factor and $\alpha$~is the angle between the propagation directions of photon and particle in the heliocentric frame~\citep{1960:Gerasimova}. Different photodisintegration reactions are possible, but by far the most likely reaction to occur, is those in which one proton or neutron is knocked out of the nucleus \citep{1993:Karakula}.

After disintegration, the charged remnants will be deflected in the interplanetary magnetic field. Since the mass\slash charge ratio will generally be different for the two fragments of the disintegrated nucleus, so will the amount of deflection be. The shape and strength of the magnetic field surrounding the Sun is quite complicated. \citet{1980:Akasofu} have constructed a three-dimensional model which consists of four components: (i) the Solar dipole magnetic field; (ii) a large number of smaller magnetic dipoles located along an equatorial circle just inside the Sun; (iii) the field of the poloidal current system generated by the Solar unipolar induction; and (iv) the field of an extensive current disc around the Sun, lying in the ecliptic plane. The influence on the deflection of cosmic particles is dominated by the latter two components, as their contributions is larger at greater distances.

Given the discreteness of the masses of the remnants and the linear proportionality between a remnant's mass and its energy (assuming single-nucleon emission), the mass number~$A$ of the original disintegrated particle can simply be determined by estimating the energies of the primaries of the two showers~\citep{1999:Epele}:
\begin{equation}
A=\frac{E_1+E_2}{E_1},
\end{equation}
where~$E_1$ is the energy of the less energetic shower.

% *****************************************************************************************************
\section{Detection of \GZ\ events}

Identifying a \GZ\ pair as such requires both showers to be seen by a cosmic ray detector. Cyclotron radii for cosmic rays at energies above $10^{16}$~eV and magnetic field strengths $≲10^{-3}$~T are very large compared to the size of the Solar system, allowing us to take both remnants' arrival directions to be equal to each other and to the original arrival direction. In order to calculate the \GZ\ detection aperture for a given cosmic ray detector, let us define the separation resulting from different amounts of deflection of the two showers as the vector~$\vec\delta=(\delta_\parallel,\delta_\perp)$ between the two tracks upon impact, transverse to the arrival direction $(\phi_0,\theta_0)$ in the Solar reference frame. Let $\delta_\parallel$ be the component in the ecliptic plane and $(\phi_0,\theta_0)=(0,0)$ in the direction of the Sun.

For an accurate description of a detector's aperture, it is necessary to incorporate the angle at which the detector is hit by the cosmic ray particles. The angles due to daily and yearly phase of Earth, $\zeta_\mathrm{d}$ and~$\zeta_\mathrm{y}$ respectively, together with the latitude~$b$ and longitude~$l$ of the detector, fix the orientation of the detector with respect to the sky. The distance between the two showers~$\vec\delta'$ depends on these four angles, and affects detection rates in a non-trivial way. As the ratio~$\delta'/\delta$ may easily exceed a factor of~$2$, it is clear that projection effects cannot be neglected in our analysis.

Whether both air showers are actually detected, depends on the detector geometry as laid out on Earth. Let us define a detector-specific function~$\xi(\vec\delta)$ which describes the probability of detecting the second shower event for a given separation vector~$\vec\delta$, under the assumption that the first shower is detected. The effective aperture~$A$ can now be calculated by integrating over~$\xi$ over the course of a year. We are also taking into account the detector's angular sensitivity~$\omega$ as a function of the zenith angle~$\theta$ as observed by the detector:
\begin{equation}\label{eq:effective-area}
	A(\vec\delta,E,\phi_0,\theta_0) = \frac{S_0}{\pi}
		\int_0^{2\pi}\int_0^{2\pi}
		\xi    (\vec\delta')
		\omega (\theta,E)
		\,d\zeta_\mathrm{d}\,d\zeta_\mathrm{y},
\end{equation}
where~$S_0$ is the total area covered by the detector, and $0\le\omega\le1$. The factor~$1/\pi$ serves to normalise to all sky visibility.

The absolute particle fluxes for various primary nuclei are estimated from the model presented by \citet{2003:Hoerandel}, which assumes
\begin{equation}\label{eq:galflux}
	J_Z(E) = J_{0,Z}
	         \left[
	             \frac{E}{E_0}
	         \right]^{\gamma_Z}
	         \left[
	             1 + \left( \frac{E}{E_pZ} \right)^{\gamma_1}
	         \right]^{-\gamma_2}.
\end{equation}
$J_Z(E)$~are the contributions of a species~$Z$ to the cosmic ray spectrum, $J_{0,Z}$~and $\gamma_Z$ are constant factors for each species, $E_0=10^{12}$~eV, $E_p=4.5\e{15}$~eV, $\gamma_1=1.9$ and $\gamma_2=1.1$. For $E≳10^{19}$~eV, this model underestimates the number of cosmic ray particles of very low mass in the spectrum. However, photodisintegration cross sections at the these energies are too low to be of consequence for very light nuclei. The total hadronic cosmic ray flux is
\begin{equation}
	J(E)=\sum_Z J_Z(E),
\end{equation}
where the summation runs over all cosmic ray particle species. In our case $2\le Z\le92$, as protons will not contribute to the \GZ\ flux at all.

If $\eta_Z(E,\phi_0,\theta_0)$ is the probability for a nucleus of species~$Z$ and energy~$E$ to undergo photodisintegration along its trajectory, then the final \GZ\ event rate for a given detector for particles with energies greater than~$E$ is given by
\begin{equation}\label{eq:rate}\begin{split}
	\Phi_\mathrm{GZ}(E)& =
		\int_E^\infty
		    \sum_Z J_Z(E')
		    \Bigg[\int\eta_Z(E',\phi_0,\theta_0)
		          A(\vec\delta,E',\phi_0,\theta_0)
		 \\ &\qquad\times
		          f_\mathrm{dc}(\phi_0,\theta_0)
		          \cos\theta_0\,d\theta_0\,d\phi_0
		    \Bigg]\,dE',
\end{split}
\end{equation}
where $f_\mathrm{dc}$ is the duty cycle of the detector, which is a constant factor in case of surface scintillators, but may depend on $\phi_0$ and~$\theta_0$ for example for air fluorescence detectors, as they cannot observe during the day.

% *****************************************************************************************************
\section{Results}\label{sec:det}
To calculate realistic values for~$\eta_Z$, a numerical model was constructed. Calculations were carried out for primary cosmic ray species from $^4$He to $^{238}$U, with energies ranging from $10^{16}$ to~$10^{20}$~eV, the region where the photodisintegration cross section is highest. The obtained average values of $\eta_Z$ for Fe, O and~He are in line with earlier findings~\citep{1999:Epele,1999:Medina-Tanco}. The partial contribution of the heavier nuclei to the \GZ\ spectrum is larger than one might expect, due to their high overall value of~$\eta_Z$.

\begin{figure}
	\includeplaatje{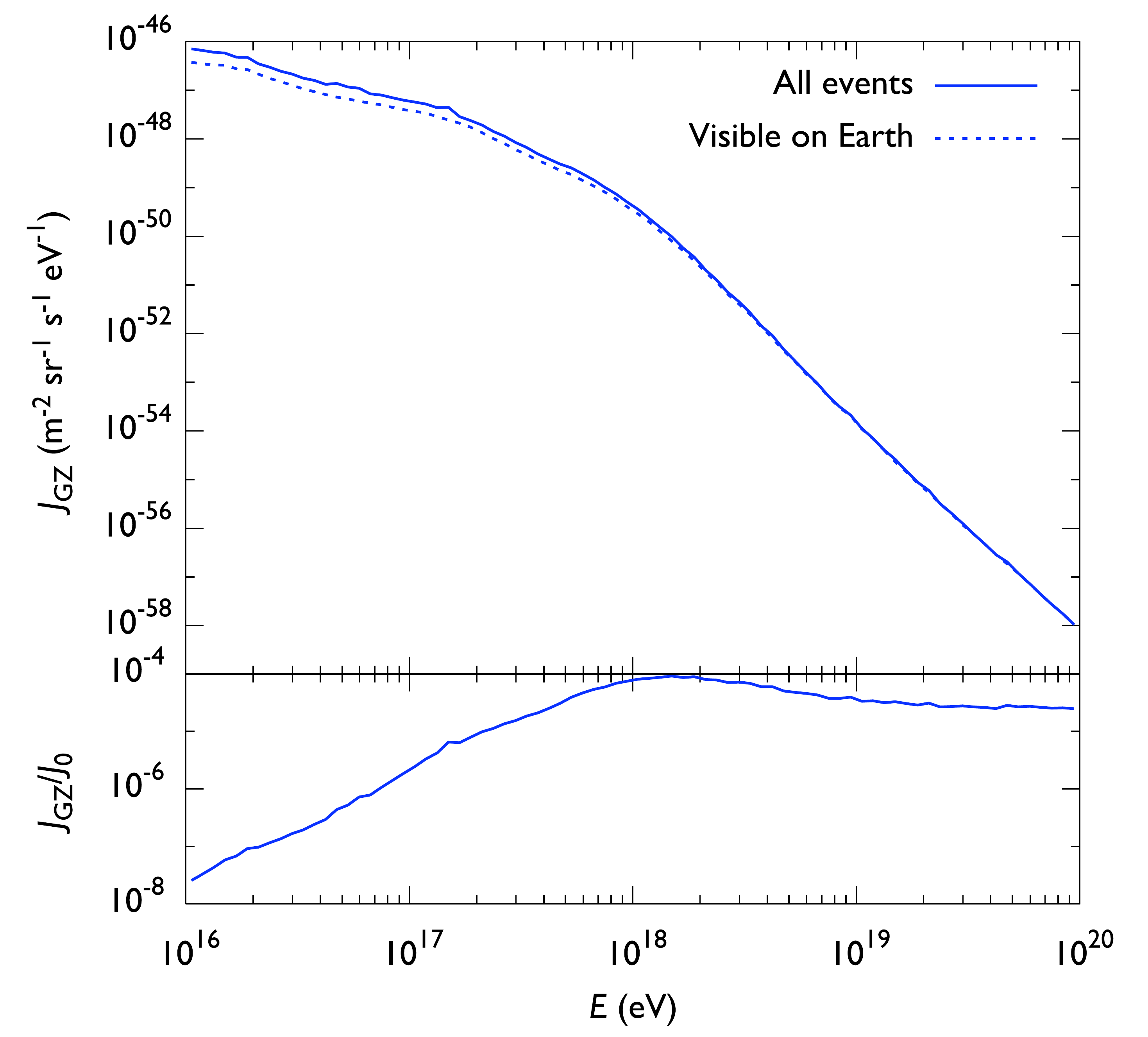}
	\caption{%
	Absolute \GZ\ energy spectrum (solid line), and upper limit for Earth-based detectors, i.e.\ events with $|\vec\delta|<2R_\oplus$ (dashed line). Also shown is the fraction of the overall galactic hadronic cosmic ray flux $J_0$  in the bottom panel, showing that $10^{-8}<\eta_\mathrm{GZ}<10^{-4}$ in this energy range.
  }\label{fig:GZflux}
\end{figure}
By multiplying each species' disintegration probability by its partial flux according to Eq.~\ref{eq:galflux}, the total intrinsic \GZ\ flux $J_\mathrm{GZ}(E)=\sum_Z J_Z(E)\int\eta_Z\cos\theta_0\,d\theta_0\,d\phi_0$ is obtained.
Fig.~\ref{fig:GZflux} shows this flux as a function of energy. The solid line represents the absolute total flux by counting all disintegration events. For reference, the flux relative to the integral cosmic ray spectrum~$J_0(E)$ is also drawn in the bottom panel, showing a maximum disintegration probability of $\eta_\mathrm{GZ}\simeq10^{-4}$ near $E\simeq1.5\e{18}$~eV. The dashed line was obtained by disregarding any event with a separation larger than one Earth diameter. This line sets a hard upper flux limit for any Earth-based detector. Notice that events with these very high separations primarily occur in the lower energy end of the spectrum: this makes sense, as the separation of a disintegrated cosmic ray pair is expected to be proportional to the inverse of its energy.

The disintegration probability strongly depends on the arrival direction. The disintegration process favours arrival directions close to the Sun, as higher integrated photon densities boost the number of disintegrations over trajectories from this direction.

Given the complexity of the magnetic field in the Solar system, a numerical equivalent of the field was implemented and disintegrated particles were propagated accordingly. Particle trajectory deviations are largest for directions near the Sun: too large, in fact, to be detected. Therefore, counterintuitively, highest event rates for any realistic cosmic particle observatory are to be expected on the night side of the sky, even though fluxes are lower there.

\begin{figure}
	\centerline{\includeplaatje{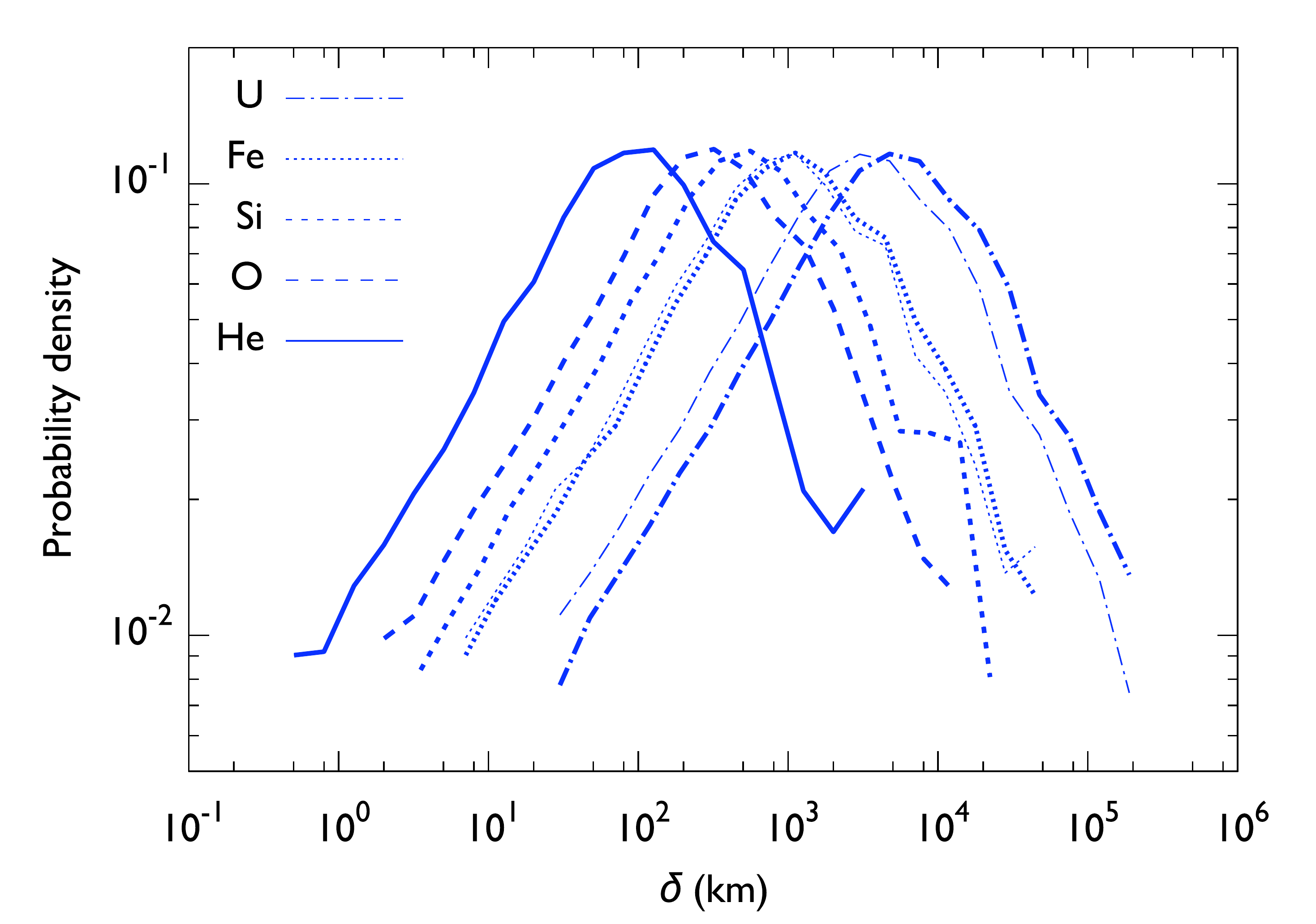}}
	\caption{%
	Probability distribution of separations for species of $^4$He, $^{16}$O, $^{28}$Si, $^{56}$Fe and~$^{238}$U. Thick lines are for proton emission, thin lines denote neutron emission. Shown are expected separations for a primary of $E=10^{18}$~eV; for other energies, multiply $\delta$ by~$10^{18}$~eV$/E$.
	}\label{fig:lambda}
\end{figure}
To illustrate the dependence of the separation distance on the various parameters, the distribution of~$\delta$ is given in Fig.~\ref{fig:lambda} for selected species. All curves are normalised so that their respective logarithmic integrals equal unity; separate lines are drawn for disintegration reactions involving proton and neutron ejection. The separations shown are for a primary of $E=10^{18}$~eV. Apart from statistical deviations, all curves are identical when separations are shifted to the left by a factor $A|Z_1/A_1-Z_2/A_2|$, peaking at $40$~km. Since $\delta\propto E^{-1}$, multiplying the separation value by $10^{18}$~eV$/E$ yields the correct separation at other~$E$. We may now parameterize the expectation value for the separation as
\begin{equation}
	\langle\delta\rangle=4A\left|\frac{Z_1}{A_1}-\frac{Z_2}{A_2}\right|\left(\frac{10^{19}~\mathrm{eV}}{E}\right)~\mathrm{km}.
\end{equation}
The overall remaining shape of the separation distribution is a result of the magnetic field shape and strength and the disintegration distance~$R$ from Earth only.

% -------------------
Let us now estimate event rates for selected air shower experiments: the existing Pierre Auger Observatory \citep{2004:Abraham} and the \lofar\ radio telescope~\citep{2006:Falcke}, which is under construction. Auger and \lofar\ are both surface detectors, sensitive to secondary shower particles (Auger) and radio signals produced in the shower (\lofar). Auger is a dense array, covering a continuous area with a total aperture of $A=4.7\e3$~km$^2$\,sr. \lofar's geometry is sparser, consisting of interconnected smaller stations with no detectors in between. Though \lofar's aperture is much smaller at $A=2.2\e2$~km$^2$\,sr, it is able to reconstruct showers with a much lower energy. Both detectors were modelled numerically to obtain accurate values for the effective aperture and \GZ\ event rate. For each detector, simulations were carried out  to make predictions for the final event rates~$\Phi_\mathrm{GZ}(E)$ according to four scenarios:
\begin{enumerate}
	\item As a simple first step, we can set a hard upper limit by taking the observatory's total aperture as effective cross section, implying $\Phi_\mathrm{GZ}(E) \simeq 0.45 A$~km$^{-2}$\,sr$^{-1}$\,yr$^{-1}$ for energies between $10^{16}$ and $10^{20}$~eV. This approach means that every event has nonzero probability of being detected, regardless of its remnants' separation.
	\item A more realistic estimate is obtained by applying the aperture function~$A$ according to Eq.~\ref{eq:effective-area}. In this way, we include projection effects as a result of the detector's orientation. We also apply a lower limit~$\delta_{\min}$ to the separation distance; this is the minimum separation at which the detector can disentangle two showers.
	\item In this scenario, a further restriction is applied by imposing a lower limit~$E_{\min}$ on the more energetic shower. For the less energetic shower, an energy down to a tenth of this limit is allowed. This approach is justified by the possible implementation of a triggering system in which data for the entire detector array is stored for each trigger, allowing one to check for coincidences at a later time.
	\item By applying a strict energy cut, demanding that both showers exceed the threshold energy, a less sophisticated trigger suffices. This scenario is a pessimistic assumption for detectors which are not optimised for \GZ\ pair detection.
\end{enumerate}

\begin{figure}
	\includeplaatje{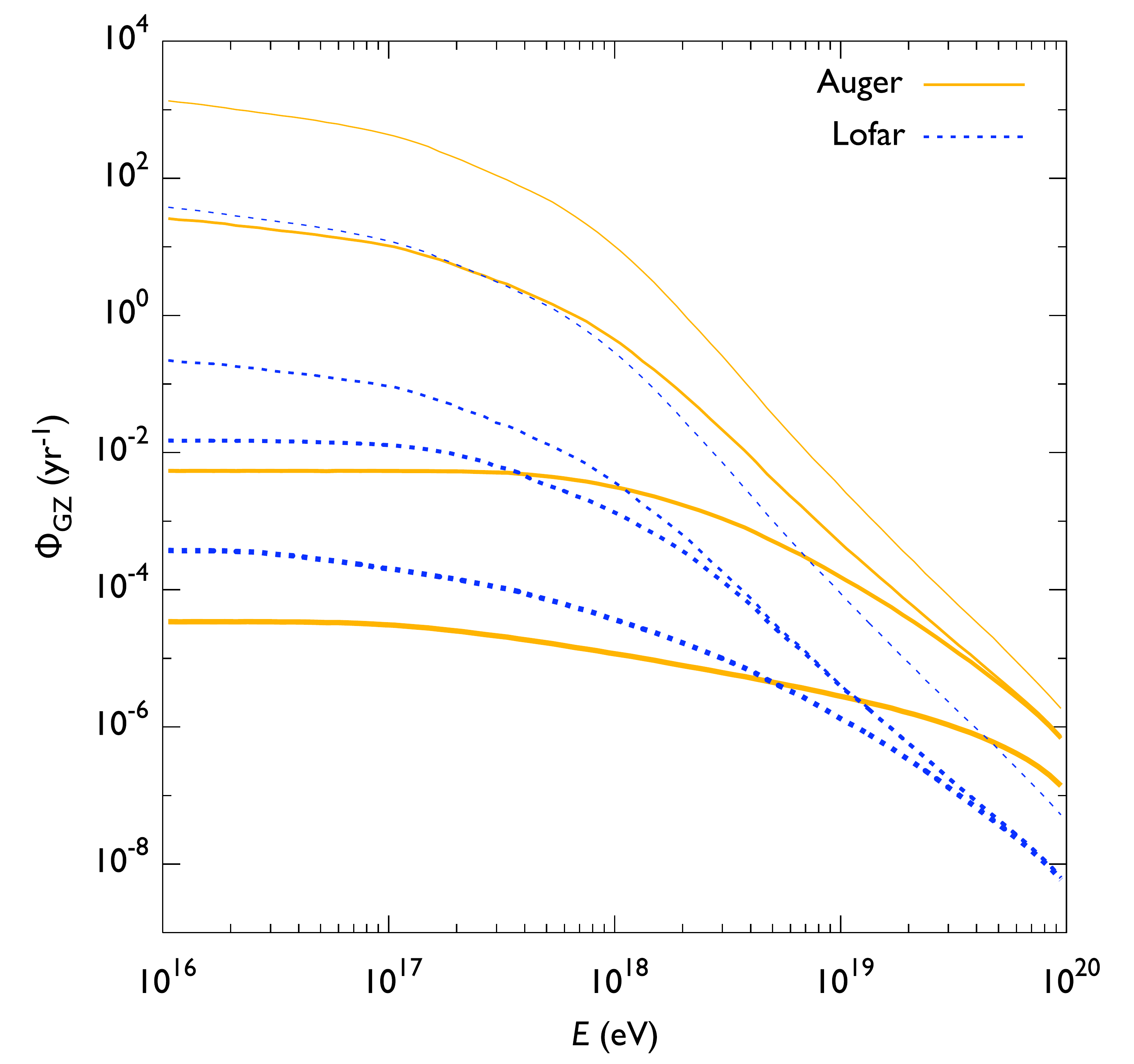}
	\caption{%
	Expected integrated \GZ\ detection rates for Pierre Auger (solid lines) and \lofar\ (dashed). Four lines are drawn for each detector, representing, from top to bottom, the theoretical upper limit for a detector of the given area; applying separation cut-offs regarding detector geometry; applying loose energy cuts; and applying strict energy cuts (see text for further explanation).
  }\label{fig:event-rate}
\end{figure}
Both detectors' $\xi(\vec\delta)$~functions were generated numerically from their geometries. The values for~$E_{\min}=10^{18}$~eV for Auger and $10^{17}$~eV for \lofar\ were interpreted as Gaussian error functions. Maximum zenith angles are $60º$ in the case of Auger and $80º$ for \lofar.

Derived event rates for each scenario are presented in Fig.~\ref{fig:event-rate}. This figure shows separate lines, from top to bottom, for each scenario. When the observatory's lower energy limit is taken into account according to scenario~4, event rates plummet to levels between $10^{-5}$ and $10^{-4}$~yr$^{-1}$. This figure effectively dismisses any possibility of successful \GZ\ pair recording. Scenario~3 is certainly a possibility in the case of Auger, as information for all surface detector tanks is stored when a big event is seen. Even in scenario~3, however, event rates are expected to stay below $10^{-2}$~yr$^{-1}$.

\lofar's detection area is about 36~times smaller than that of Auger, and consequently it has a much lower intrinsic \GZ\ flux, not exceeding $0.3$~yr$^{-1}$. \lofar's energy limit is 10~times lower, however: this easily compensates the lack of area, as the slope of the \GZ\ event rate is approximately $\propto E^{-3}$ at higher energies (see Fig.~\ref{fig:event-rate}). Still, expected event rates do not exceed $5\e{-4}$~yr$^{-1}$ in scenario~4. Comparing event rates from simulations of the full detector and the central core only, shows that no significant contribution to pair detection is to be expected from the relatively small outer stations. The increased rate in scenario~3, probably the maximum achievable rate for \lofar, is still quite insignificant at $2\e{-2}$~yr$^{-1}$. The chances of the corresponding trigger algorithm being implemented are slim at best, because the \lofar\ telescope will not be a dedicated cosmic ray detector, but an experiment shared with other astronomical observations. In practice, data bandwidth limitations would probably not allow a trigger to be communicated to every antenna in the array, except for very energetic events which produce a negligible \GZ\ rate.

\begin{figure}
	\includeplaatje{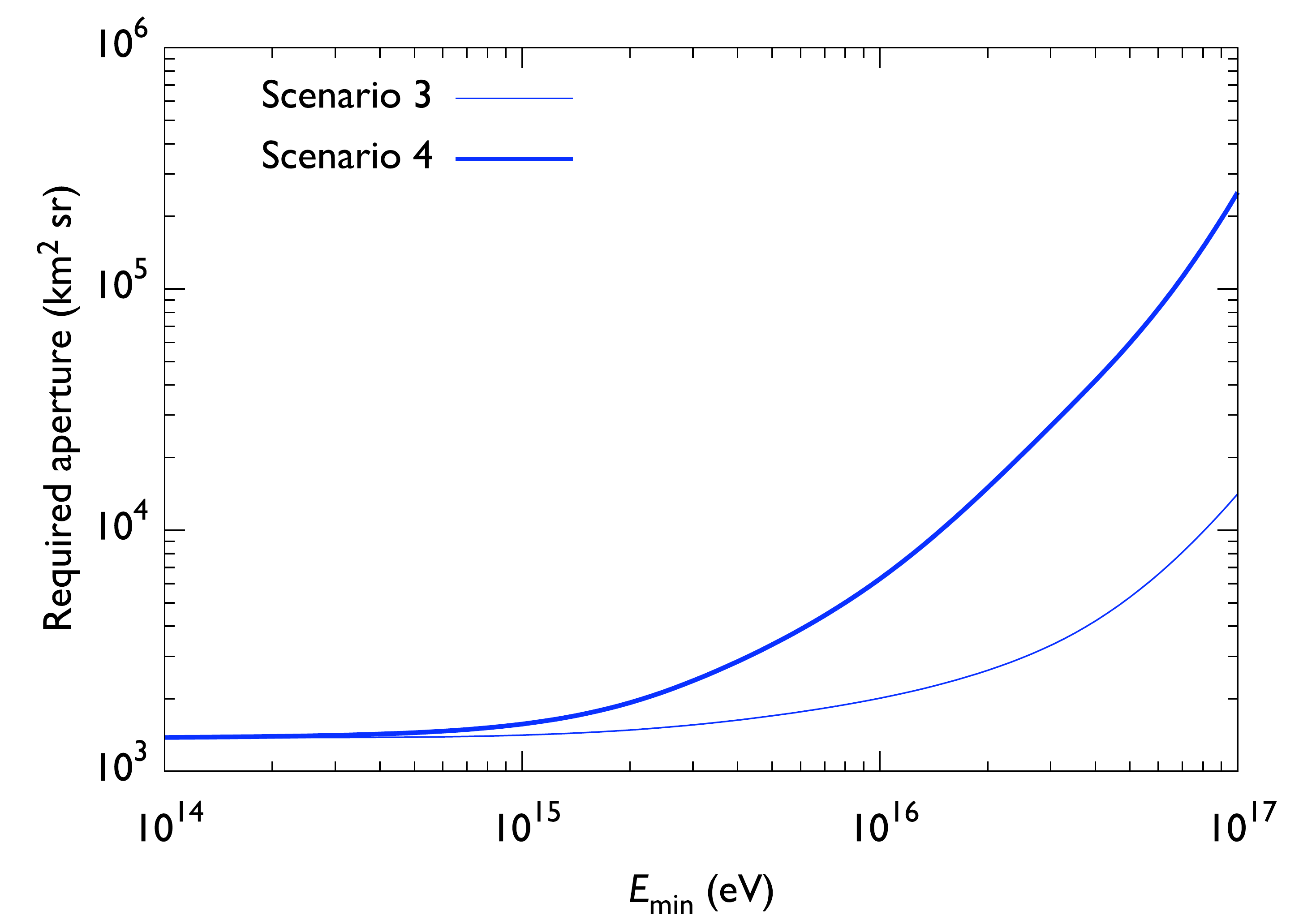}
	\caption{%
	Required aperture for detection of a single \GZ\ event per year as a function of $E_{\min}$ for a circular detector at $45º$~latitude, $f_\mathrm{dc}=1$, $\theta_{\min}=60º$ and $\delta_{\min}=1$~km.
  }\label{fig:required-aperture}
\end{figure}
Because the expected event rate for a cosmic ray detector is approximately proportional to $S_0E^{-3}$, it pays off to invest in lowering a detector's energy threshold instead of focusing on collecting area if maximising \GZ\ event rates is intended. Still, any detector would have to be huge to receive one single detectable event per year or more. Estimated required aperture sizes are shown in Fig.~\ref{fig:required-aperture} for a detector with a Gaussian lower energy limit~$E_{\min}$ with full duty cycle and $60º$~zenith limit. At Auger's energy limit, the required aperture in scenario~3 would be $\sim90$~times that of Auger; at \lofar's energy limit, it would be $\sim50$~times that of \lofar. Even at low threshold energies, where an aperture of only $\sim1.4\e3$~km$^2$\,sr would be required, the cost of constructing and operating such an array would easily amount to several billions of euros.

For surface detectors, there might be other points to consider. Even if a cosmic ray observatory was hit by a detectable \GZ\ pair, it is very much the question whether such an event would meet the quality criteria. The occurrence of two overlapping or nearly overlapping showers within a very short time window might prevent proper reconstruction, and the event would be discarded as noise.

Fluorescence detectors will not produce more favourable \GZ\ detection rates, as their duty cycles are typically only~$0.1$, increasing the aperture needed by an order of magnitude over surface detectors. Moreover, the fluorescence detection technique does not work well for low energies, where rates are highest: typically, a $10^{17}$~eV air shower's fluorescence signal can be seen no more than $10$~km away from a fluorescence telescope. This also discards satellite missions such as \textsc{euso} as possible detectors.

On a side note, an interesting deviation of the pure cosmic ray spectrum as seen by current experiments can be observed. We have shown that a fraction of $\sim10^{-5}$ of all cosmic particles that arrives at Earth does so as a \GZ\ pair. In practice, these events go unnoticed as such and are registered as normal cosmic particles. This means that, at any given energy, a small fraction of the events is to be attributed to a primary of much higher energy. The actual spectrum then has a spectral index of the order of~$10^{-5}$ less steep than currently assumed; this factor is too small to perceive given current data error margins.

% *****************************************************************************************************
\section{Conclusion}\label{sec:conc}
We have used a set of simulations to calculate in detail the rate of very high energy cosmic ray particles breaking up in the Solar magnetic field and the probability distributions of the separations of their remnants. Additionally, we have used accurate detector models to estimate realistic detection rates for this phenomenon. We have shown that current experimental setups, including the Pierre Auger Observatory, by far lack either the energy sensitivity or the area to produce any significant amount of detections of the kind, and would detect only a fraction of the \GZ\ flux predicted by \citet{1999:Epele} and \citet{1999:Medina-Tanco}. Consequently, the prospects for any future experiment detecting the \GZ\ effect are negligible.

%  *****************************************************************************************************
\acknowledgements{This work is part of the research programme of the `Stichting voor Fundamenteel Onderzoek der Materie (\textsc{fom})', which is financially supported by the `Nederlandse Organisatie voor Wetenschappelijk Onderzoek (\textsc{nwo})'.}

\bibliographystyle{aa}

\end{document}